\documentclass[12pt]{amsart}
\usepackage{latexsym}
\usepackage{amsmath}
\usepackage{epsfig}
\usepackage{amsfonts}
\usepackage{amssymb}
\usepackage{amsthm}
\usepackage{graphicx}
\DeclareGraphicsExtensions{.jpg}

\newtheorem{theorem}{Theorem}
\newtheorem{lemma}[theorem]{Lemma}
\newtheorem{proposition}[theorem]{Proposition}

\theoremstyle{definition}
\newtheorem{definition}[theorem]{Definition}

\begin{document}

\title{Ropelength of tight polygonal knots}

\author{Justyna Baranska, Piotr Pieranski}
\address{Poznan University of Technology \\
Laboratory of Computational Physics and Semiconductors \\ 
Nieszawska 13A, 60 965 Poznan, Poland \\
E-mail: pieransk@man.poznan.pl, jmarch@phys.put.poznan.pl
}

\author{Eric J. Rawdon}

\address{Department of Mathematics and Computer Science\\
Duquesne University\\ 
Pittsburgh, PA 15282, USA\\
E-mail: rawdon@mathcs.duq.edu}

\begin{abstract}
A physical interpretation of the rope simulated by the SONO algorithm
is presented. Properties of the tight polygonal knots delivered by the
algorithm are analyzed.  An algorithm for bounding the ropelength of a
smooth inscribed knot is shown. Two ways of calculating the ropelength
of tight polygonal knots are compared. An analytical calculation
performed for a model knot shows that an appropriately weighted
average should provide a good estimation of the minimum ropelength for
relatively small numbers of edges.
\end{abstract}

\maketitle 

\section{Introduction}

Knots tied on material objects, e.g.{} a rope of a finite thickness,
are called physical knots \cite{SimonVegas}. It is well known from
everyday life that open knots can be tightened by pulling apart the
loose ends of the rope. Tightening a closed knot, i.e.{} a knot tied
on a rope whose ends are spliced, is more difficult since one has to
engage processes in which the rope will reduce its length while
keeping its diameter intact. Such a process, difficult to create in
the laboratory, can be easily simulated numerically and we find in the
literature various examples of such simulations
\cite{SKK,physicalknotsbook,dissertation,spatial,mecancomputers,janathesis,maddocksbiarcs}.
From the intuitive point of view, it is rather obvious that in
tightening a knot tied on a rope of a finite diameter, one must arrive
at a conformation at which the tightening process stops. This
conformation is called \textit{tight}. Any
small change in the shape of a tight conformation needs an elongation
of the rope. Obviously, tight conformations of knots tied on different
types of rope, with different physical properties, will be
different. Thus, to make the notion of the tight knot unambiguous, one
must specify the physical model of the rope on which the knot is
tied. The simplest of such models is the \textit{perfect
rope}. Perfect rope is \textit{perfectly floppy} -- needs no force to
be bent, but, at the same time, it is \textit{perfectly hard} -- needs
an infinite force to be squeezed, its diameter always remains
intact. It is also \textit{perfectly slippery}. Its formal definition
will be given in the next section.

Tight conformations can be seen as minimizers of the ropelength
function \cite{bosimple,lsdr}. The minima can be only
local. While it has been shown that global minima
exist \cite{cksminimum,olexistence}, with the exception of the
trivial knot and some simple links, the minimizing conformations are
not known. Knots in the conformations at which the ropelength reaches
its global minimum are called {\it ideal}. As long as the ideal
conformations are not known analytically, we cannot be
certain that the conformation at which one arrives is ideal.

In all simulations of the processes in which knots are tightened, the
knots are represented by a finite sequence of points in $\mathbb{R}^3$
space. When connected, the points can be seen as vertices of a
self-avoiding polygon. Most often the simulation programs either start
from a non-equilateral conformation and aim at making it equilateral
(e.g.{ }SONO) or start from an equilateral conformation and try to
keep it such (crankshaft rotations). Let $P$ be such an equilateral
polygon. The essential problem we face is to construct, using the
coordinates of the vertices of $P$, a smooth knot $K_P$ which can be
seen as the axis of the knot tied on the perfect rope. In constructing
$K_P$, we make some assumptions concerning the shape of the
pieces of which it is built. In what follows we assume that
the knot $K_P$ is built from smoothly connected arcs inscribed into
$P$, which makes the knot $C^{1,1}$ smooth. As a result of this assumption
and provided the radius of the rope is properly chosen, we show 
that the knot $K_P$ can indeed be tied with perfect rope
leaving it self-avoiding. On the other hand, we also know that the
inscribed knot is not ideal, thus, its ropelength can serve only as an
upper bound for the minimum ropelength of the unknown ideal
conformation. In finding tight polygonal knots $P$ with an increasing
number of vertices, we see that the ropelength of the inscribed smooth
knots $K_P$ converges to a value, which we believe is the ropelength
of the ideal conformation, although we will not prove that here.  This
convergence is slow however, and one might want an approximation using
relatively few edges.  This is especially important for complicated
knots where the number of edges needed to adequately approximate the
smooth ropelength would not be practical.  With this aim, a second,
weighted-average, approach is explored.  This technique provides
surprisingly accurate approximations with relatively few edges.
Together, we provide both a theoretically-sound approach with slow
convergence and a practical approach with fast convergence.  Depending
on the application, we expect that both techniques will be useful.

\section{Computing upper bounds for smooth knots inscribed in tight polygons}

\subsection{Ropelength of knots tied on the perfect rope}

Smooth ropelength models perfect rope as a non self-intersecting tube with
a $C^{1,1}$ knot as its core.
For completeness, we include these
definitions and some properties of this knot energy.

\begin{definition}
For a $C^{1}$ knot $K$ and $x\in K$, let $D_r(x)$ be the disk
of radius $r$ centered at $x$ lying in the plane normal to the tangent
vector at $x$.  Let 
$$R(K)=\sup\{r>0\,:\,D_r(x)\cap D_r(y)=\emptyset
\text{ for all }x\not = y \in K\}\,.$$
The quantity $R(K)$ is called the 
\textit{thickness radius} of $K$.
Define the \textit{ropelength} of $K$ to be 
$$Rope(K)=Length(K)/R(K)\,,$$
where $Length(K)$ is the
arclength of $K$.
\end{definition}

The thickness radius $R(K)$ is the radius of a thickest tube that can
be placed about the knotted core $K$ without self-intersection.  Note
that in some of the literature, the thickness refers to the
\textit{diameter} of this thickest tube, in which case the ropelength
is half of what we use here. The problem we face is how, knowing the
shape of $K$, to find $R(K)$.

Suppose we are given a concrete knot $K$ and we are trying to
reproduce it with an impenetrable tube. Thus, $K$ serves here as the
core of the knotted tube. We have to pay attention to two types of
interactions between the tube's normal disks that restrict the radius
$R$ of the tube.

\begin{enumerate}
\item[(A)] The interactions between the discs centered on pairs of points
located infinitesimally close to the diagonal of $K \times K$ -- these
normal disks will start overlapping when the curvature
radius is smaller than $R$.
\item[(B)] The interactions between the discs centered on points located away
from the diagonal of $K \times K$ -- the disks will start overlap when the
Euclidean distance between the points is smaller than $2R$.
\end{enumerate}

\begin{figure}
  \centering
    \includegraphics[height=4.5in]{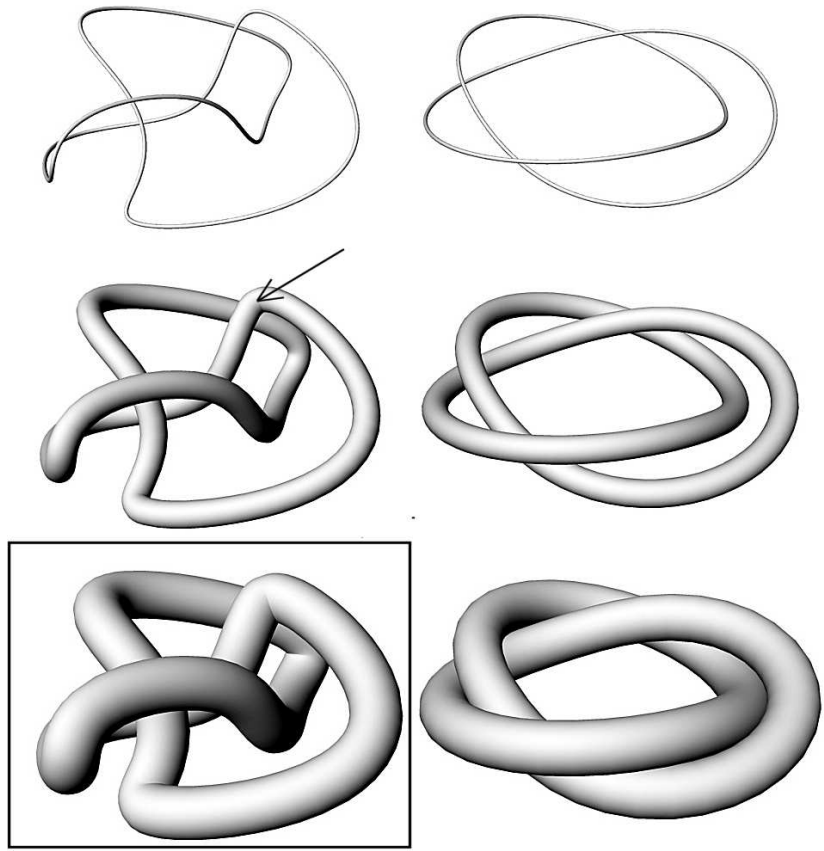}
  \caption{Two different conformations of the trefoil knot are inflated. 
  Both conformations have the same arclength. The inflation of the
  conformation shown on the left stops first because of its sharp
  bends at which interactions of type A intervene. Further inflation would
  lead to the development of singularities in the tube surface.
  See the framed picture. Inflation of the conformation shown
  on the right stops at the point above which interactions of type B
  intervene.}
  \label{fig:1}
\end{figure}

Figure \ref{fig:1} illustrates the problem.  These intuitive
observations are captured by the quantities below and the subsequent
lemma.

\begin{definition}
For a $C^2$ knot $K$ with unit tangent map $T$, 
let $MinRad(K)$ be the minimum of the radius of curvature
measured at each of the points on the knot.  
The \textit{doubly-critical self-distance} is the minimum distance 
between pairs of points on the knot
whose chord is perpendicular to the tangent vectors
at the both of the points.  
In other words, let
\begin{equation*}
DC(K)=\{(x,y)\in K\times K\,:\,T(x)\perp\overline{xy}\perp T(y), 
x\not = y\}
\end{equation*}
where $\overline{xy}$ is the chord connecting $x$ and $y$.
Define the \textit{doubly-critical self-distance} by
\begin{equation*}
DCSD(K)=\min\{|x-y|\,:\,(x,y)\in DC(K)\},
\end{equation*}
where $| \cdot |$ is the standard $\mathbb{R}^3$ norm.
\end{definition}

There is a fundamental relationship between $R(K)$, $MinRad(K)$, and
$DCSD(K)$.

\begin{lemma} Suppose $K$ is a $C^2$ knot. Then 
$$R(K)=\min\left\{MinRad(K),\frac{DCSD(K)}{2}\right\}\,.$$
\label{thicknessc2}
\end{lemma}
\begin{proof}
See \cite{lsdr}.
\end{proof}

In \cite{cksminimum,olexistence}, it is shown that ropelength minima exist as 
(at worst) $C^{1,1}$ curves.

After reworking the definition of $MinRad$,
one can extend Lemma \ref{thicknessc2} to include
$C^1$ (and thus $C^{1,1}$ curves).
The following is taken from \cite{lsdr2}.
For a $C^0$ function
$f:\mathbb{R}\to \mathbb{R}^3$,
define the \textit{dilation} of $f$ by 
\begin{equation*}
dil(f)=\sup\left\{
\frac{| f(s)-f(t)|}{|s-t|}\,:\,s,t\in \mathbb{R}, s\not = t\right\}.
\end{equation*}
Note that if $K$ is a $C^2$ knot parameterized by
arclength with unit tangent map 
$T$, then
$MinRad(K)=1/dil(T)$.  Thus, the dilation gives a generalization for
$MinRad$ to knots that are $C^1$.  Since 
$MinRad$ and $1/dil(T)$ are equal for
$C^2$ knots, we use $MinRad$ to denote $1/dil(T)$ for all $C^1$ knots.
If a knot is $C^1$ but not $C^{1,1}$, then $dil(T)=\infty$, 
in which case $MinRad$ is assumed to be $0$.  
In this paper, we are mainly interested in $C^{1,1}$ knots, 
in which case $dil(T)$ is
finite and $MinRad$ is positive.
Alternate, but equivalent, approaches 
for defining the ropelength of $C^{1,1}$ knots are explored in 
\cite{maddocks,cksminimum}.

\begin{lemma}
Suppose $K$ is a $C^1$ knot.  Then
$$R(K)=\min\left\{MinRad(K),\frac{DCSD(K)}{2}\right\}\,.$$  
Furthermore,
if $K$ is $C^{1,1}$, then $R(K)>0$.
\label{thickness}
\end{lemma}
\begin{proof}
See \cite{cksminimum} or \cite{lsdr2}.
\end{proof}

\subsection{Ropelength of polygonal knots}

As mentioned in the introduction, the problem we face in analyzing knots
provided by numerical simulations is that they are not given
in a smooth continuous form. For instance, programs based on the SONO
algorithm deliver equilateral polygonal knots. Let $P$ be such a
knot. It is represented by a sequence of $n$ points whose positions
are indicated by vectors $\vec{v}_1, \vec{v}_2, \ldots , \vec{v}_n$. We
shall refer to them as the \textit{vertices} or \textit{beads}
of the polygonal knot $P$
that one obtains by joining consecutive vertices.  We define

\begin{align*}
\vec{s}_i &= \vec{v}_{i+1} - \vec{v}_i\\ 
s_{i} &=\left|\vec{s}_i\right|\\
\vec{r}_{i,j} &= \vec{v}_{j} - \vec{v}_i\\ 
r_{i,j} &= \left|\vec{r}_{i,j}\right|\,.
\end{align*}

The vectors $\vec{s}_i$ will be called \textit{edges}.  When referring to
the vertices as points on a polygonal knot, we shall omit the
vector notation.

\noindent \textbf{Remark:} Because the knots that we are considering
are closed, whenever the vertex index $k$ happens to be larger than
$n$, we assume that $\vec{v}_k = \vec{v}_{k-n}$. Similarly, when the
index happens to be smaller than 1, we assume $\vec{v}_k =
\vec{v}_{k+n}$.

In what follows, we will use the notion of the index distance
$ID$ between the vertices:

\begin{equation*}
 ID(v_i,v_j) = \left\{
 \begin{array}{ll}
 \left|i-j\right|, & \text{when} \left|i-j\right|\leq{n/2} \\
 n-\left|i-j\right|, & \text{otherwise}
  \end{array}
  \right.\,.
\end{equation*}

Note that alternate approaches have been appeared in
\cite{stasiak1,stasiakideal,meideal,ksthickness,mine,devrthickness2,maddocks,mecancomputers,janathesis,maddocksbiarcs}.
In the SONO simulations, the vertices of a knot interact with each
other as if they were spheres, the properties of which will be later.
Taking into account the representation, we shall extend the notion of
the thickness radius in terms of the virtual spheres.  We do this by
defining notions of $MinRad$ and $DCSD$ for polygons similar to the
characterization theorem (Lemma \ref{thickness}) for smooth thickness
radius.

\begin{figure}
  \centering
    \includegraphics[height=2.6in]{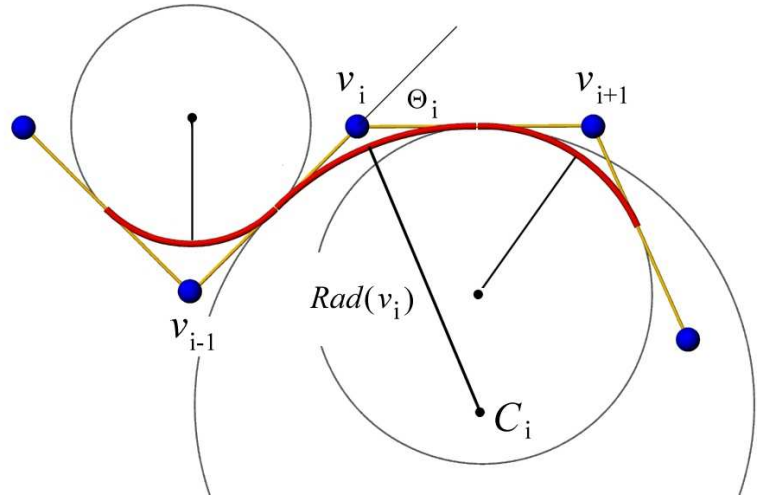}
  \caption{Vertices and edges of a piece of an equilateral polygonal
  knot $P$. Inscribed arcs of radii $Rad(v_i)$ are also shown. To
  simplify the drawing we assumed that all of the vertices are located
  in the same plane. }
  \label{fig:2}
\end{figure}

\begin{definition}
Suppose $P$ is an equilateral polygonal knot with edge lengths $dl$.
Let
$$Rad(v_i) = \frac{dl}{2\tan(\theta_i/2)}$$
where $\theta_i$ is the turning angle of the tangent vectors at $v_i$.
Note that $Rad(v_i)$ is the radius of an arc of a circle $\alpha$ 
that can be inscribed in the corner at $v_i$ so that $\alpha$ intersects both
adjacent edges at the midpoint and the tangent at the midpoints coincide
with the tangents on $\alpha$ at the midpoints.  Let
$$MinRad(P)= \min_i\{Rad(v_i)\} =
\frac{dl}{2\tan(\theta_{max}/2)}\,,$$
where $\theta_{max}$ is the maximum turning angle. See Figure \ref{fig:2}.
\end{definition}

\begin{definition}
Let $P$ be a polygonal knot.  We define the {\it sphere thickness
radius} to be
$$R_s(P) = \min\{\sqrt{MinRad(P)^2+(dl/2)^2}, SR(P)\}$$ where $SR(P)$, the {\it sphere
radius}, is half the minimum distance between vertices whose index distance
is at least $\lceil \pi MinRad(P)/dl-1\rceil$.
\end{definition}

The construction of the sphere thickness radius needs a few words of
explanation.  Suppose $P$ is an equilateral knot. Imagine that about
each of the vertices, there is a sphere of radius $R$. When $R$ is
larger than $dl/2$, the spheres about nearby vertices
intersect. The cases in which $dl/2<R<dl/\sqrt{2}$ need special attention, 
but since they never occur in the simulations we eventually analyze, 
we exclude them assuming implicitly in what follows that $R>dl/\sqrt{2}$. 
The surface of the union of all spheres is bumpy, so we
refer to this as the {\it corrugated tube} or {\it bead rope}
about the knot.  The problem we face is to inflate the spheres
as much as possible without violating the condition of the
self-avoidedness of the corrugated tube formed by their union.

\begin{figure}
  \centering
    \includegraphics[height=1.6in]{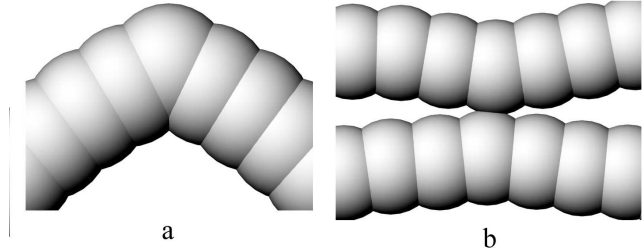}
  \caption{Violation of the self-avoidedness of the bead tube. (a) The
  tube bends too quickly. (b) Two index-distant portions of the tube
  are too close to each other. }
  \label{fig:3}
\end{figure}

First, let us explain what we mean by self-avoiding in the case of
the corrugated tube. Consider three consecutive vertices $v_{i-1}$,
$v_i$, and $v_{i+1}$ and the spheres $S_{i-1}$, $S_i$, and $S_{i+1}$,
around them all of radius $R$. The spheres $S_{i-1}$ and
$S_i$ intersect at a circle $C_{i-1}$ whose radius equals
$R_D=\sqrt{R^2-(dl/2)^2}$. The spheres $S_i$ and $S_{i+1}$ intersect at a
circle $C_{i}$ of the same radius. Let $D_{i-1}$ and $D_i$ be the
disks filling the circles. The disks are inclined at an angle equal to
the turning angle $\theta_i$. We assume that the disks are hard and
hence cannot overlap; thus, at most they may become tangent at a single
point. The latter happens when the sphere radius $R_i=\frac{dl}{2
\sin(\theta_i/2)}$. For a given polygonal knot $P$,
the smallest of such radii is
$\sqrt{MinRad(P)^2+(dl/2)^2}$. When $R$ is chosen larger
than $\sqrt{MinRad(P)^2+(dl/2)^2}$, we see the behavior shown 
in Figure \ref{fig:3}a.

When $R\leq \sqrt{MinRad(P)^2+(dl/2)^2}$, all portions of the
corrugated tube are well-defined locally and one must check if index-distant
vertices are too Euclidean-close. Let us explain, what we mean by
stating that the portions of the corrugated tube are
well-defined. Each portion of the corrugated tube is a sphere with two
pieces cut off. The cutting planes are perpendicular to the
polygon edges and are located at the midpoints. Consecutive
portions are thus connected with disks of radius $R_D$.  As stated
above, the disks are assumed to be hard and they are not allowed to
overlap.  A portion of the corrugated tube which fulfills the
condition is said to be well-defined.  If the Euclidean distance
between two vertices is smaller than $2R$, the portions of the
corrugated tube would overlap.  Thus, we check all the Euclidean
distances between sufficiently index-distant vertices and find the
smallest one. This will be the value of $2 SR(P)$. This is a
geometrical interpretation of the self-avoidedness conditions found in
the definition of $R_s(P)$. When $R$ is chosen larger then $SR(P)$, we
see the behavior shown in Figure \ref{fig:3}b.

We will be mainly interested in optimizing $R_s$, but
to do so we need a scale-invariant version of $R_s$. There are
various ways to do this and we explore two normalizations in the next
section.  The corrugation will play a part in how the knot packs
efficiently in these optimizing conformations. In the short term, our
goal is to show that for a sufficiently ``thick'' polygon $P$, there
exists a smooth knot inscribed in $P$ whose thickness radius is close
to $R_s(P)$.  We first present the inscribing algorithm.  A similar
formulation for inscribing polygons appears in \cite{mecancomputers}.

\begin{figure}
  \centering
    \includegraphics[height=2.7in]{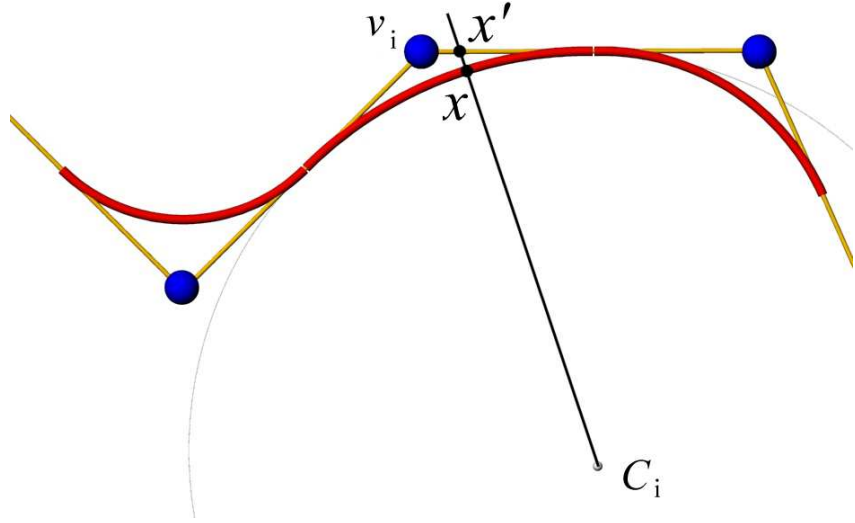}
  \caption{The definition of the points $x$ and $x'$ used in
  Proposition \ref{howtoinscribe}.}
  \label{fig:4}
\end{figure}

\begin{proposition}
For a given $n$-edge equilateral polygonal knot $P$ with edge length $dl$,
a $C^{1,1}$ curve $K_P$ 
can be inscribed in $P$ in such a way that
$MinRad(K_P) = MinRad(P)$ .  Furthermore, 
there exists a bijection between $K_P$ and $P$ so that for each
pair $x'\in P$ and $x\in K_P$, we have
$| x-x'| \leq R_s(P)\left(\sec 
\left(\frac{\theta_{max}}{2}\right)-1\right)$.
\label{howtoinscribe}
\end{proposition}
\begin{proof}
An arc $\alpha_i$ of a circle of radius $Rad(v_i)$
can be inscribed at $v_i$
such that $\alpha_i$ is tangent to $\vec{s}_{i-1}$ and $\vec{s}_i$ 
and intersects
the adjacent edges at the midpoints.
Let $K_P$ be the result of inscribing arcs of radius $Rad(v_i)$ 
at each vertex $v_i$ and removing the bypassed corners. 
Since there is no overlapping of adjacent inscribed arcs,
$K_P$ is well-defined as a (possibly self-intersecting) closed curve.
The curve $K_P$ has a piece-wise constant radius of curvature, and
$MinRad(K_P) = MinRad(P)$.
The knot $K_P$ is $C^1$ and piecewise $C^2$, and thus, 
lies in the category of $C^{1,1}$ curves.  

For each $x$ on the inscribed curve $K_P$, we define a unique point $x'$
on $P$.  If $x$ is a midpoint on $P$, let $x'=x$.  
Otherwise, $x$ lies on an arc, say $\alpha_i$, whose center is $C_i$.  
Let $x'$ be the intersection of the ray 
$\overrightarrow{C_ix}$ with $e_{i-1}\cup e_i$ (see Figure \ref{fig:4}).
Simple trigonometric calculations show that 

\begin{equation*}
\Vert x-x'\Vert \leq MinRad(P)\left(
\sec\left(\frac{\theta_{max}}{2}\right)-1\right).
\end{equation*}
\end{proof}

For each $x\in K_P$, we say the \textit{corresponding vertex} is the 
vertex which generates the arc on which $x$ lies.
The midpoint of each edge has two corresponding
vertices.  It will be convenient to think of the midpoint as corresponding
to each vertex in different situations and this will not pose any problems
in this work. Note that the smaller $\theta_{max}$ is, the
closer the inscribed curve $K_P$ will be to the polygon $P$.  

\begin{figure}
  \centering
  \includegraphics[width=5.5in]{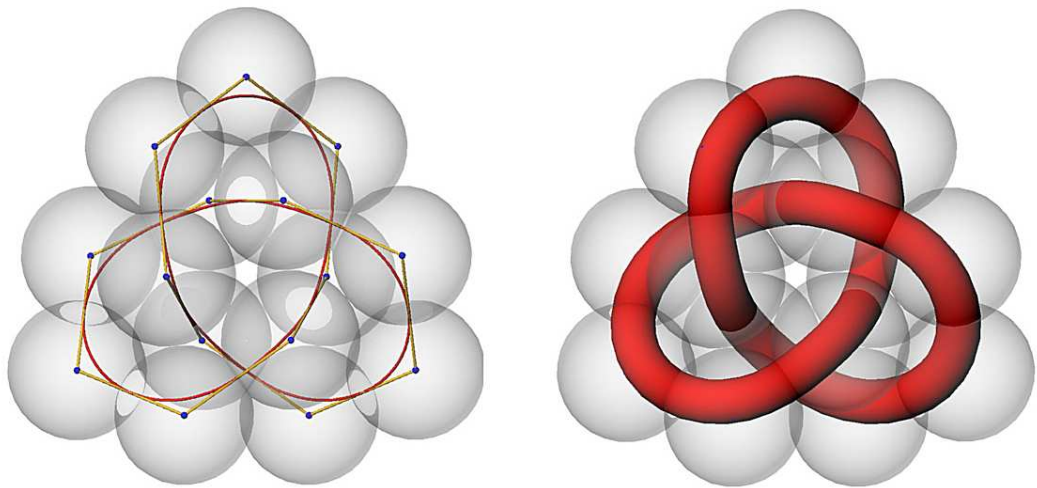}
  \caption{The picture on the left presents the polygonal trefoil knot
    $P$ tightened by SONO. The distance between its vertices is guarded by
    the virtual spheres of radius $SR(P)$. Since in the case shown in the
    figure $SR(P)<\sqrt{MinRad(P)^2+(dl/2)^2}$, the value of $SR(P)$
    serves as the sphere thickness radius $R_s(P)$. Consecutive spheres
    meet at circles of radius $R_c(P) = \sqrt{R_s(P)^2-dl^2/4}$. The
    smooth knot $K_P$ shown also in the figure consists of arcs inscribed
    into $K$. The picture on the right shows the inscribed knot $K_P$
    inflated to radius $R_c(P)$. It is well visible, and we prove it in
    the text, that the smooth tube knot, hidden as a whole inside the
    corrugated tube made of the virtual spheres, remains self-avoiding.}
  \label{fig:5}
\end{figure}

The corrugated tube about the polygon is a union of many spheres, some of which
are allowed to intersect. The intersection between the surface of two adjacent spheres consists of a
circle whose radius is $\sqrt{R_s(P)^2-dl^2/4}$ and whose center
is the midpoint of the edge.  This value will be important, so
we define $R_c(P) = \sqrt{R_s(P)^2-dl^2/4}$.
We will show now that the smooth tube of radius $R_c(P)$ about $K_P$ is non-self
intersecting, i.e.{} that the thickness radius $R(K_P)$ of $K_P$ is at least
$R_c(P)$. See Figure \ref{fig:5}. This is a bit delicate and will take some
effort to prove.  We will need the following technical lemma.

\begin{lemma}
Let $P$ be an equilateral polygonal knot with edge lengths
$dl$ and $R_s(P)>dl/\sqrt{2}$ 
and $K_P$ be the smooth 
knot inscribed in $P$
via the algorithm of Proposition \ref{howtoinscribe}.  
If $x,y\in K_P$ and $arc(x,y)\geq \pi MinRad(K_P) = \pi MinRad(P)$,
then $ID(v_x,v_y) \geq \lceil \pi MinRad(P)/dl - 1\rceil$, 
where $v_x$ and $v_y$ are the 
vertices corresponding to $x$ and $y$.
\label{round}
\end{lemma}
\begin{proof}
Suppose $arc(x,y)\geq \pi MinRad(P)$.  Now $arc(x,y)$ is the minimum arclength
on $K_P$ measured over the two paths connecting $x$ and $y$.  By the construction of the inscribed smooth knot, we know that $arc(x,y)$ is
no larger than the arclength along $P$ between the corresponding
points $x'$ and $y'$.  In the worst case, both $x$ and $y$ are midpoints
and their corresponding vertices $v_x$ and $v_y$ lie in the shorter arc
between $x$ and $y$.  In such a case, we have
$$arc(x,y) \leq arc_P(v_x,v_y) + 2(dl/2)\,,$$
where $arc_P$ denotes the minimum arclength when walking along the
polygon $P$.
Now $arc_P(v_x,v_y) = n \,dl$ where $n$ is the number of edges between
$v_x$ and $v_y$.  Thus,
$n\,dl \geq \pi MinRad(P) - dl$ or $n\geq \pi MinRad(P)/dl - 1$.
\end{proof}

In other words, if $arc(x,y)\geq \pi MinRad(P)$, then we know that the
corresponding vertices $v_x$ and $v_y$ will be a part of the set over
which $SR(P)$ is computed.  This will be important in ensuring that
$SR(P)$ bounds $DCSD(K_P)$.

% The following lemma (for a proof, see \cite{cancomputers})
% is also needed for the proof of the theorem.  Roughly, it states that
% the $DCSD$ must be realized on a pair of points whose arc distance
% is at least $\pi MinRad$.
% 
% \begin{lemma}
% Let $A(K)=\{(x,y)\in K\times K\,:\,arc(x,y)\geq \pi MinRad(K)\}$.
% For a $C^{1,1}$ knot $K$, 
% $$R(K)=\min\left\{MinRad(K),\min_{(x,y)\in A(K)}
% \frac{| x-y|}{2}\right\}\,.$$
% \label{arcchar}
% \end{lemma}

We now have the tools to bound the thickness radius of the inscribed
knot $K_P$.

\begin{theorem}
Let $P$ be an equilateral polygonal knot with edge lengths $dl$ and
$R_s(P)>dl/\sqrt{2}$
and $K_P$ be the smooth 
knot inscribed in $P$
via the algorithm of Proposition \ref{howtoinscribe}.  
Then
$R(K_P) \geq R_c(P) = \sqrt{R_s(P)^2 -\frac{dl^2}{4}}$\,.
\end{theorem}
\begin{proof}
By the construction of $K_P$, we know that
$MinRad(K_P) = MinRad(P)$.
We split the proof into two cases: when 
$DCSD(K_P)/2 \geq MinRad(K_P)$ and when $DCSD(K_P)/2 < MinRad(K_P)$.

In the former case, we have
$R(K_P) = MinRad(K_P) = MinRad(P) \geq R_c(P)$.

In the latter case, we assume $R(K_P) = DCSD(K_P)/2$.  By
\cite{mecancomputers}, we know that when $DCSD(K_P)$ is realized at a
pair of points, say $a$ and $b$, then $arc(a,b) \geq \pi MinRad(P)$.
Therefore, the corresponding vertices $v_a$ and $v_b$ have
$ID(v_a,v_b) \geq \lceil \pi MinRad(P)/dl-1\rceil$ by Lemma
\ref{round}.  Let $D_a$ and $D_b$ be the disks of radius $R_c(P)$
normal to $K_P$ at $a$ and $b$ respectively and let $B_a$ and $B_b$ be
the spheres of radius $R_s(P)$ centered at $v_a$ and $v_b$
respectively.  Now $a$ and $b$ lie inside $B_a$ and $B_b$
respectively, and by the definition of $R_s(P)$, we know that the
interiors of $B_a$ and $B_b$ do not intersect.  Furthermore, the
normal disks $D_a$ and $D_b$ are contained within $B_a$ and $B_b$
respectively.  Therefore, $D_a$ and $D_b$ can only intersect, in the
worst case, on the boundary.  This implies that $|a-b|\geq 2 R_c(P)$
or $DCSD(K_P)/2 \geq R_c(P)$.
\end{proof}

The previous result is not sharp in general.  However, for the sake of
this paper, it does provide a computable lower bound for the thickness
radius of $K_P$.  Furthermore, the length of $K_P$, which is smaller
than the length of $P$, can be computed explicitly.  We call
$Length(K_P)/R_c(P)$ the \textit{inscribed ropelength} and denote it
$L_c$.  The value of $L_c$ is an upper bound for the ropelength of the
inscribed $K_P$, and as $dl\to 0$ (note that changing $dl$ affects
$n$, $P$, and $K_P$), our bound on $R(K_P)$ approaches $R_s(P)$.
Since $R_s(P)>dl/\sqrt{2}$, the inscribed arcs lie within the
corrugated tube about $P$, which guarantees that $K_P$ has the same
knot type as $P$.  Since $K_P$ is a $C^{1,1}$ smooth knot with the
same knot type as $P$, we have a bound on the ropelength of one smooth
knot within the knot type.  Thus, we have an upper bound for the
minimum smooth ropelength for the given knot type.

\section{Interpretation of simulations performed with the SONO algorithm}

Let us now describe the SONO algorithm using the notions defined
above. First notice that in discussing the ropelength of smooth knots,
we are considering knots $K$ of a given length $Length(K)$ looking for
their thickness radius $R(K)$. The value $R(K)$ is the maximum radius
for which $K$ could be inflated without violating the conditions that
the surface of the tube must remain self-avoiding. The value
$Rope(K)=Length(K)/R(K)$ is the ropelength of $K$. The application of
the virtual inflation suggests numerical calculations in which the
inflation process forces the knot to change its conformation. Such
changes may maximize $R(K)$ and thus minimize $Rope(K)$. Working with
polygonal knots, it proves more convenient to consider a different
scheme in the simulations: keep the sphere radius $SR$ fixed and
shorten the edge length $dl$ so that the knot eventually arrives at a
conformation for which $R_s(P)=SR$. This is the approach we discuss
below.

\subsection{Basic procedures of SONO}

There are two essential procedures on which SONO is based
\cite{pierideal}. The basic goal of SONO is to reduce
$Length(P)$ subject to the constraint that $R_s(P)\geq
SR$, for some fixed rope radius $SR$. In practice, we choose $SR=1$. 
The reduction of the polygon length is achieved by
reducing its edge length $dl$.  We aim at simulating equilateral
knots; thus one of these procedures called EqualizeEdges (EE), checks
the length of the edges, and, if they differ from the desired value
$dl$, introduces necessary corrections. Suppose the distance $s_i$
between vertex $i$ and $i+1$ is different from the desired $dl$
value. Then the vertices are shifted toward new positions so that
their distance is closer to $dl$:

\begin{equation*}
\begin{array}{ll}
\vec{v'}_{i} = \vec{v}_{i}-c_{1} (dl-s_{i})\:\vec{s}_{i}/s_{i}\\
\vec{v'}_{i+1} = \vec{v}_{i+1}+c_{1} (dl-s_{i})\:\vec{s}_{i}/s_{i}
\end{array}
\end{equation*}
$c_1\in(0,1/2]$. Usually we work with $c_1=1/2$.

As a result of multiple applications of the procedure, the dispersion
of the edge lengths in the final conformations becomes very small, and
thus, we consider the knots we get from our simulations as equilateral.

The second procedure called RemoveOverlaps (RO), checks the distances
between the vertices. When the RO procedure finds that the minimum
distance $2SR$ is violated, the vertices $v_i$ and $v_j$ are shifted
away from each other to a distance equal to $2SR$ or, what proves to
speed up the initial stage of the tightening process, exceeding $2SR$
by $\epsilon$:

\begin{equation*}
\begin{array}{ll}
\vec{v'}_{i} = \vec{v}_{i}-c_2
(2SR-r_{i,j}+\epsilon)\:\vec{r}_{i,j}/r_{i,j}\\ \vec{v'}_{j} =
\vec{v}_{j}+c_2 (2SR-r_{i,j}+\epsilon)\:\vec{r}_{i,j}/r_{i,j}
\end{array}
\end{equation*}
where $c_1\in(0,1/2]$ and is usually set equal to 1/2. The value of $\epsilon$
changes during the tightening process. Initially it is of the order
$10^{-2}$ and reaches the level of $10^{-7}$ at its end. 

The procedures RO and EE are in some circumstances contradictory, but
their multiple application leads to simultaneous reduction of both the
overlaps and the dispersion of the edge lengths. Both parameters are
constantly monitored.

When the edge length $dl$ is small, the tightened polygonal knots
develop short regions where $Rad(v_i)$ tends to become small. An
additional procedure ControlCurvature (CC) monitors this, never
allowing the external angle between the consecutive edges to be larger
than $2 \arcsin \frac{dl}{2SR}$. This, in particular in the final
stages of the tightening, makes sure that the tight conformation of
$P$ will have $R_s(P)=SR$.

\subsection{Physical sense of the SONO algorithm and practical details of simulations}

Let us now discuss the physical sense of the simulation based on the
SONO algorithm. The simulated knot $P$ can be seen as tied on a closed
necklace of beads. The necklace is unusual, because its beads are
unusual: if their index distance $ID(v_i,v_j) \leq \lceil \frac{\pi
MinRad(P)}{dl} \rceil$, they are allowed to overlap, otherwise they
behave as hard spheres of radius $SR$ and repel infinitely hard. The
RO procedure simulates this interaction. Because of the action of the
EE procedure, the centers of the consecutive beads can be seen as
connected with nonextensible rods of controlled length $dl$. The rods
are connected to each other by elasticity free but angle-limiting
joints. Each of the beads has the shape of a sphere with two parts of
it cut off.  The cutting planes are located in the middle of the
edges which lead to it.  Consecutive beads are thus connected via
disks of radius $R_c=\sqrt{SR^2-dl^2/4}$.  The necklace can be bent in
any direction but only to some extent: the CC procedure limits the
angle between the consecutive rods to $ 2 \arcsin \frac{dl}{2SR}$. As
a result, $MinRad(P) \geq R_c$, which prevents the disks separating
consecutive beads to overlap.  At most they are allowed to become
tangent.  Because of its construction, the surface of the bead rope is
not smooth. It is corrugated and the corrugation is more pronounced,
the larger the distance between consecutive beads. This distance is, of
course, the edge length $dl$.

The tightening process runs as follows. Suppose we start with a loose
conformation. The EE and RO procedures make the lengths of all
edges equal and remove overlaps between the beads. After multiple
applications of the procedures, the dispersion of the edge lengths
and the sum of all overlaps fall below a fixed level.  Then SONO reduces the
required edge length $dl$. Now, as the EE procedure tries to adjust
the length of the edges to the new value by shortening the distance
between consecutive beads, new overlaps may appear. Subsequently they
are removed by the RO procedure. And so on. 

In the final tight conformation, the dispersion of edge lengths and
the sum of all overlaps are smaller than $10^{-6}$ which allows us to
consider the closest beads as just touching. Keeping in mind that the
CC procedure does not allow $\sqrt{MinRad(P)+dl^2/4}$ to be smaller
than $SR$, we may consider the final tight knot as a polygonal,
equilateral knot $P$ whose $R_s(P)=SR$.

The shrinking rope forces the knot to change its conformation. At the
end, we arrive at a conformation for which further shortening is no
longer possible because it creates non-removable overlaps. Obviously,
if the number of beads is small, the final value of $dl$ is large and
the bead rope is strongly corrugated. This may create problems, since
the knot is more likely to become stuck in a local minimum.  The
reason for this is as follows: one part of the rope winding around
another part may get into a groove between consecutive beads even if
shifting it to another groove would allow further shortening of the
rope.  The shifting will not be achieved by SONO, because it would
need a (rather large) temporary increase of the ropelength. Working
with a small number of beads, we set $\epsilon$ to a higher value to
minimize these effects. In general, the tightening process should be
seen as a physical experiment in which the experimenter watches and
adjusts the parameters.

\subsection{The problem of finding the right ropelength, an experimental approach}

\begin{figure}
  \centering
  \includegraphics[scale=0.33]{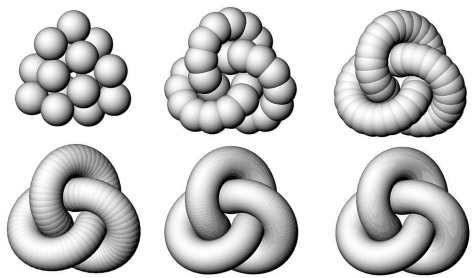}
  \caption{Tight conformations of the trefoil knot tied on a
    rope consisting of 15, 30, 60, 120, 240 and 480 beads}
  \label{Trefoil 15-480}
\end{figure}

In Figure \ref{Trefoil 15-480}, we have the tightest conformations of
the trefoil knot found by SONO working with rope consisting of $n=$15,
30, 60, 120, 240, and 480 vertices.

\begin{figure}
  \centering
  \includegraphics[scale=0.50]{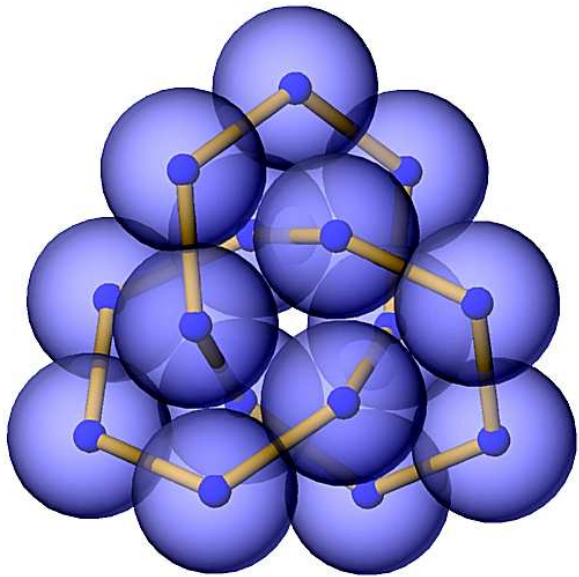}
  \caption{Details of a trefoil knot tightened by SONO. The knot
    is tied on a piece of corrugated rope consisting of $n=15$
    beads. The centers of the beads make the vertices of the polygonal
    knot. The vertices are connected with straight cylindrical
    edges.}
  \label{15mx}
\end{figure}

It seems plausible that the best estimation of the minimal length of the
rope needed to tie a particular knot will come from analyzing the polygonal
conformation obtained for highest $n$. As seen in the figure, at $n=480$ the
surface of the bead rope is visually smooth -- the corrugation of its
surface, so visible at $n=15$, $30$, $60$, becomes
undetectable by the naked eye. The problem we face is that in tightening
much larger knots, for instance the $(2,99)$ torus knot, reaching this
level of the surface smoothness would require a very large
number of beads, which is both awkward and time-consuming. For practical
reasons, it is reasonable to limit the number of beads as much as
possible. We present a picture showing the details of the SONO
tightened trefoil for such a small $n$ in Figure \ref{15mx}.

Calculations run faster when $n$ is small, but what about the accuracy
of the ropelength we can extract from the analysis of the final,
tight conformation? 

\section{Ropelength of SONO knots}

We are mainly interested in understanding the optimal conformations
of $R_s$.  However, $R_s$ varies with scale, so one could always increase
$R_s$ simply by scaling the polygon.  In the second section, we explored
one way to normalize $R_s$, namely by analyzing the ropelength $L_c$ of
the inscribed knot $K_P$.  In doing so, we overestimate the minimum
ropelength.  In this section, we present a different normalization
that will underestimate the ropelength.  In the next section,
we combine these two notions to obtain a reasonable approximation of
the minimum ropelength with relatively few edges.

The simplest way to determine the ropelength is to sum the
length of all edges of the knot and divide by $R_s(P)$:

\begin{equation*}
L_{p}=\left(\sum_{i=1}^{n}s_i\right)/R_s(P) = n\,dl/R_s(P)\,.
\end{equation*}

In what follows we shall refer to $L_p$ as the {\it raw polygonal 
ropelength}. In calculating the ropelength in such a simplistic manner, we
implicitly assume that the polygonal knot makes the axis of a rope of
radius 1, which is obviously wrong. Certainly, if the cylindrical
segments of the rope are not to overlap, the radius should be smaller.
This becomes clear for large $dl$. However erratic the raw polygonal
length may seem, it will prove to be very useful.

\begin{figure}
  \centering
  \includegraphics[height=3in]{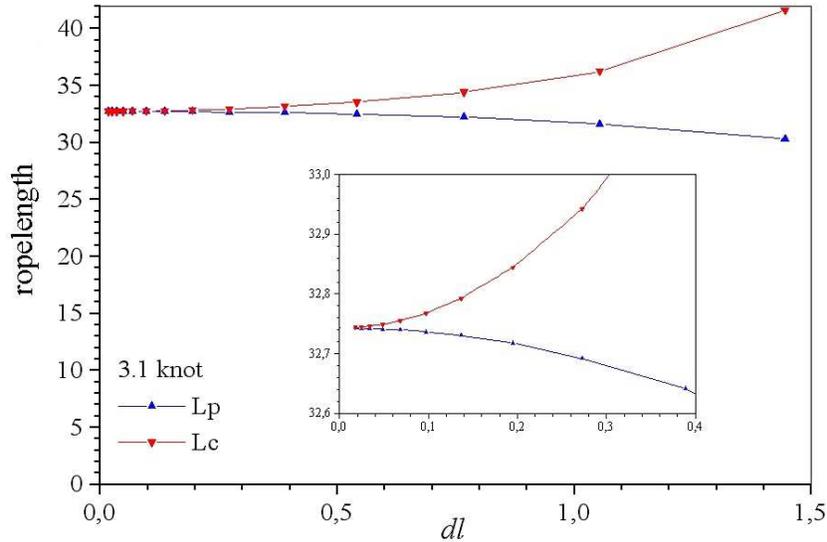}
  \caption{The raw polygonal $L_p$ and inscribed $L_c$ ropelength
    for SONO tightened trefoil knot versus the edge
    length with $n$=21, 30, 42, 60, 84, 120, 168, 240, 336, 480, 672,
    960, 1342 and 1920.}
  \label{31LpLc}
\end{figure}

It is interesting to see how the values of the raw polygonal and
inscribed ropelength behave in practice as $n$ increases. For the sake
of convenience, Figure \ref{31LpLc} shows both values plotted versus
the edge length $dl$.

First notice that as the edge length diminishes, the
raw polygonal and the inscribed ropelengths apparently converge to a
common value.  We call this value the \textit{true ropelength},
denoted $L_{\infty}$.  In spite of its erratic
nature, the raw polygonal ropelength converges faster than the
inscribed ropelength.  The raw polygonal ropelength $L_p$ 
underestimates while the
inscribed ropelength $L_c$ overestimates the ropelength. Can one 
find a weighted average that will
accurately estimate the ropelength with relatively few edges? 
Below, we present a heuristic
reasoning which provides the answer.

\subsection{The problem of finding the right ropelength, an 
analytic approach}

\begin{figure}
  \centering
  \includegraphics[scale=0.50]{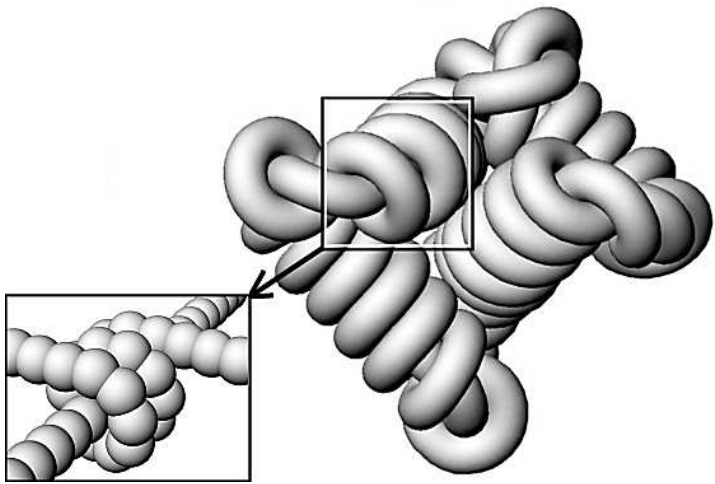}
  \caption{A tight conformation of the (2,59) torus knot found by
    SONO. A typical situation where one piece of the rope is
    winding around another piece is shown in the
    frame. In simulations, the rope is corrugated as shown in the
    inset.}
  \label{TypicalSituation}
\end{figure}

By looking at SONO tightened knots, see for instance Figure
\ref{TypicalSituation}, one may notice that we often deal with pieces
of a rope winding around other pieces of the rope. At smaller $n$, the
rope is strongly corrugated.  Thus, to be precise, we deal with pieces
of a corrugated rope winding around other pieces of the corrugated
rope. In a model situation shown also in the picture, the central
piece is straight and the other piece winds tightly around
it. By analyzing the raw polygonal and inscribed ropelength functions in
this simplified model, we will shed some light on their functioning in
the description of tight knots.

\begin{figure}
  \centering
  \includegraphics[scale=0.50]{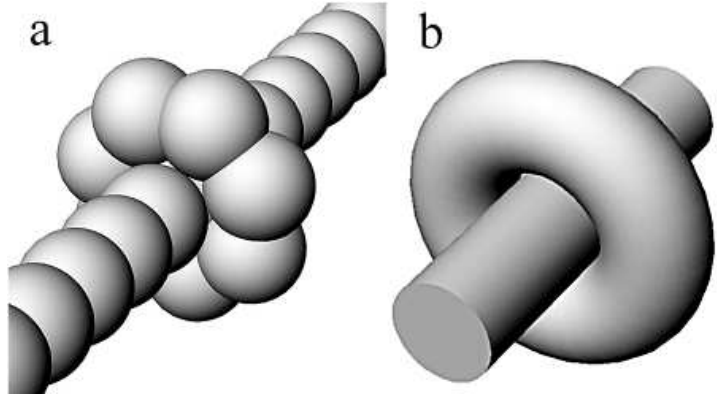}
  \caption{(a) A piece of a corrugated rope winds toroidally
    around the corrugated cylinder (a straight piece of corrugated
    rope). (b) In the $n \rightarrow \infty$ limit, we deal with a
    torus winding around a cylinder.}
  \label{ToroidalModel}
\end{figure}

To analyze the problem, we consider a still simpler case of a
corrugated rope wound toroidally around a corrugated cylinder. See
Figure \ref{ToroidalModel}.

In the latter case, we may perform a rigorous analysis of the ways in
which the raw polygonal and the inscribed ropelength estimate the true
value of the ropelength. The latter is known since it is simply the
length of a smooth torus winding tightly a smooth cylinder.  Assuming
that both the cylinder and the torus are made of a tube whose radius
$R=1$, the value is $4\pi$.

Now, let us consider the corrugated torus on the corrugated
cylinder. Both of them can be seen as unions of spheres. 
We assume that all the spheres have radius $R=1$.
The distance between consecutive spheres is $dl$. 
Obviously, for an arbitrary $dl$ there are problems with a
clean closing of the torus. In what follows we shall concentrate
on a short piece of the torus, so this is not problematic.

The best way to wind a piece of a corrugated rope around a straight
piece of the same rope is to wind it inside the groove between two
consecutive spheres. Figure \ref{ToroidalModel} illustrates the
situation. Our aim is to find the raw polygonal length and the
inscribed ropelength of a single piece of the rope and compare them
with the true length of the analogous piece of the torus wound
tightly around the cylinder. The relations we find here should apply
to the situation we face in calculating the ropelength of the SONO
tightened knots.

\begin{figure}
  \centering
  \includegraphics[height=1.5in]{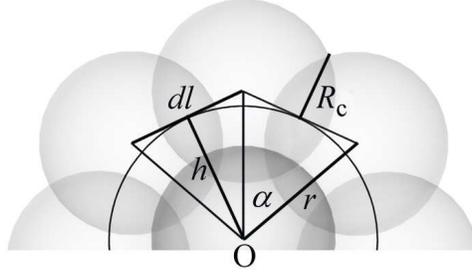}
  \caption{The geometry of the corrugated rope wound within a groove of a
    straight piece of the same rope. The value of $R_c$ is the radius of the
    circles at which consecutive spheres intersect. Thus, it is also
    the radius of the smooth rope which can be safely placed inside
    the corrugated rope. Its axis runs along the inscribed arcs.}
  \label{ModelGeometry}
\end{figure}

Let us look at Figure \ref{ModelGeometry}. Assume the distance between
the spheres, i.e.{} the polygon edge length, is $dl$. Since the
corrugated rope is wound within a groove, its spheres are at a
distance $r<2$ from the center $O$ of the axis of the corrugated
cylinder.  We get
 \begin{equation*}
r=\sqrt {4-dl^2/4}\,.
\end{equation*}

The distance is different when the corrugated rope is wound not within the
groove, but on the hill of the corrugated cylinder. Here $r=2$. In
real situations, such as this presented in Figure
\ref{TypicalSituation}, the corrugated rope runs both in the grooves and on
the hills. It is thus reasonable to assume an average $r$. It would be
an error to calculate the average as an arithmetic average, since the
shape of the hills is not saw-tooth. The hills have the form of
circular arcs.

\begin{figure}
  \centering
  \includegraphics[scale=0.25]{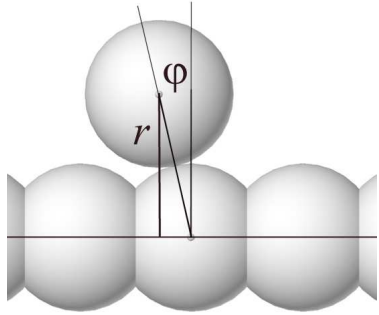}
  \caption{The bead of the winding rope can be located anywhere
    on the circular hill between two consecutive grooves. }
  \label{Averaging_r}
\end{figure}

The position of the bead can be parametrized by the angle $\varphi$,
whose maximum $\varphi_{max}$ value is given by
\begin{equation*}
\varphi_{max}=\arcsin \left(\frac{dl}{4}\right)\,.
\end{equation*}
See Figure \ref{Averaging_r}. The average value of $r$ is found by
integration:
\begin{equation*}
	r_{av}=\frac{1}{\varphi_{max}} \int^{\varphi_{max}}_{0} {2
	cos(\varphi) d \varphi}=\frac{dl}{2 \arcsin (\frac{dl}{4})}\,.
\end{equation*}

Let us return now to Figure \ref{ModelGeometry}. The angle $\alpha$ at
which the edge is seen from the center $O$ is given by
\begin{equation*}
	\alpha = 2\arcsin\left(\arcsin\left(\frac{dl}{4}\right)\right)\,.
\end{equation*}

The middle point of the edge is found at the distance $h$ from $O$,
where
\begin{equation*}
	h=\frac{dl\,{\sqrt{{\arcsin (\frac{dl}{4})}^{-2}-1}}}{2}\,.
\end{equation*}

The length of the inscribed arc joining the middle points of two
consecutive edges is
\begin{equation*}
d\lambda=dl\,\arcsin (\arcsin (\frac{dl}{4}))\,{\sqrt{{\arcsin
\left(\frac{dl}{4}\right)}^{-2}-1}}\,.
\end{equation*}

Since the arc makes the axis of the smooth rope of radius
\begin{equation*}
R_c=\sqrt{1-dl^2/4}\,,
\end{equation*} 
the normalized length $dL_{c}=d\Lambda / R_c$ of the rope
segment is
\begin{equation*}
  dL_c=\frac{2\,dl\,\arcsin \left(\arcsin \left(\frac{dl}{4}
\right)\right)\, {\sqrt{ {\arcsin (\frac{dl}{4})}^{-2}-1}}}{{\sqrt{4 -
{dl}^2}}}\,.
\end{equation*}

The raw polygonal length of the segment is simply
\begin{equation*}
	dL_p=dl\,.
\end{equation*}

Now, let us ask the basic question: what is the true ropelength $dL$
of the segment? It is the length of a piece of a smooth torus (wound
tightly on the cylinder of unit radius) seen at angle $\alpha$ from
the center $O$:

\begin{equation*}
  dL=4\arcsin \left(\arcsin \left(\frac{dl}{4}\right)\right)\,.
\end{equation*}

\begin{figure}
  \centering
  \includegraphics[scale=0.34]{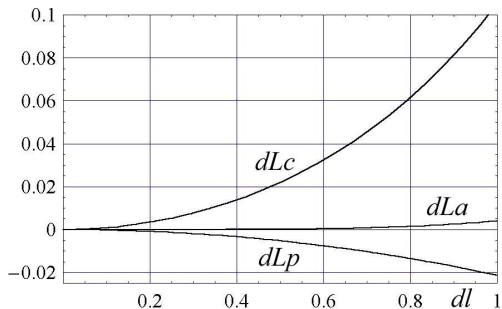}
  \caption{The relative deviations of the raw polygonal $dL_p$
    and inscribed $dL_c$ segment lengths from the true value
    $dL$. The value $dL_a$ is the relative deviation of the
    weighted average $dL_a=(4 dL_p+dL_c)/5$. Compare the plots
    with the plots shown in Figure \ref{31Data}.}
  \label{dLcompare}
\end{figure}

The relationship between the true length $dL$, its raw polygonal 
length $dL_p$,
and its inscribed length $dL_c$ estimations should be similar to those
observed for the true $L_\infty$, raw polygonal $L_p$, and inscribed
$L_c$ ropelengths found for knots tightened by SONO. Thus, let us
see how the raw polygonal $dL_p$ and inscribed rope $dL_c$
approximations differ from the true value $dL$. To get a clear
quantitative estimation of errors that we make using $dL_p$ and
$dL_c$, we plot the relative deviations $(dL_p-dL)/dL$ and
$(dL_c-dL)/dL$ of the raw polygonal and inscribed segment length from
its true $dL$ value. See Figure \ref{dLcompare}.

By looking at the picture, one can clearly see that $dL_p$ underestimates
while $dL_c$ overestimates the length of the segment.  Perhaps 
an appropriately weighted average of the values could
provide a better estimation of the true length. To check this,
we solve the equation
\begin{equation*}
	(a dL_p + dL_c)/(a+1)=dL\,.
\end{equation*}

Its solution is
\begin{equation*}
  a=\frac{2\,\left( 2 - \frac{dl\,{\sqrt{{\arcsin (\frac{
    dl}{4})}^{-2}-1}}}{{ \sqrt{4 - {dl}^2}}} \right) \,\arcsin
    (\arcsin (\frac{dl}{4}))}{dl - 4\,\arcsin (\arcsin
    (\frac{dl}{4}))}\,.
\end{equation*}

\begin{figure}
  \centering
  \includegraphics[scale=0.34]{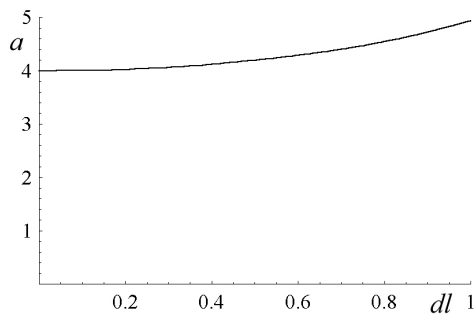}
  \caption{Weight $a$ versus $dl$.}
  \label{a_versus_dl}
\end{figure}

The functional dependence of the weight $a$ on the edge length $dl$
looks rather complex, but its plot versus $dl$ reveals that the
dependence is very weak -- at $dl=1$, its value is close to 5, but as
$dl$ diminishes, it converges quickly to 4. See Figure \ref{a_versus_dl}.

By looking at the plot, one may suspect that the expansion of $a(dl)$
should have zero order term equal to 4 and much smaller higher order
terms. We carried out the expansion to get:
\begin{equation*}
	a=4 + \frac{353\,{{dl}}^2}{480} + {O({dl}^4)}\,.
\label{expansion}
\end{equation*}

\begin{figure}
  \centering
  \includegraphics[height=3.6in]{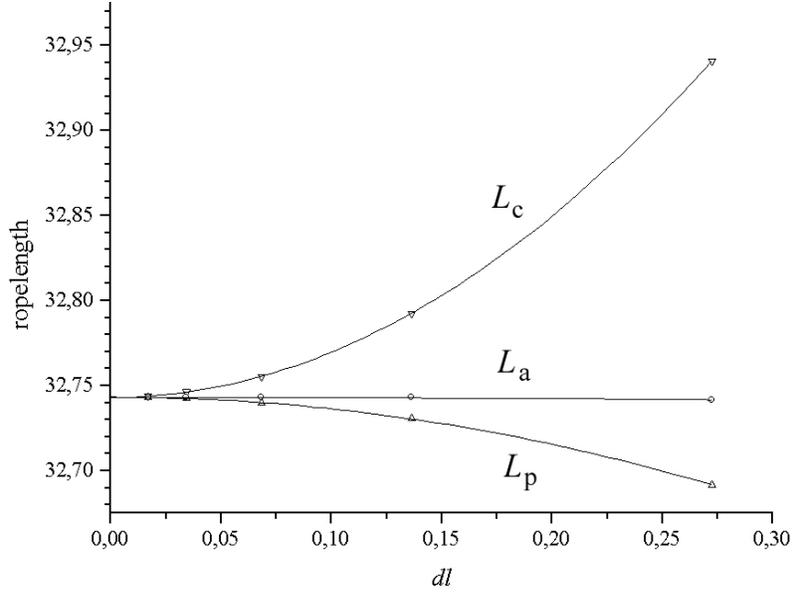}
  \caption{Raw polygonal $L_p$, inscribed $L_c$, and weighted
    average ropelength $L_a$ for SONO tightened trefoil knot versus the
    edge length for $n$= 240, 480, 960 and
    1920. The data were fitted with with second order polynomial curves.}
  \label{31Data}
\end{figure}

Since the relations between $L_p$, $L_c$ and $L_{\infty}$ in tight knots
should be similar to the relations between $dL_p$, $dL_c$ and $dL$, we
arrive to the conclusion that the weighted average
\begin{equation*}
	L_a=(4 L_p + L_c)/5
\end{equation*}
should be almost independent of the edge length $dl$ and, thus, it
should provide a good estimate of the $L_{\infty}$ value. To check the
hypothesis, we perform a series of tests. First of all, we calculate
the weighted average for the trefoil knot, whose $L_p$ and $L_c$ are
plotted in Figure \ref{31Data}. Looking at the plot of $L_a$, one
clearly sees that, as expected, it is almost independent of $dl$.

\begin{table}
  % \centering
  \caption{The raw polygonal, inscribed and weighted average
    ropelength of the trefoil knot tightened by SONO.}
    {\begin{tabular}{|c|c|c|c|c|} 
      \hline
      $n$ & $dl$ & $L_p$ & $L_c$ &$L_a=(4 L_p+L_c)/5$ \\
      \hline 
      120  &	0.27243068 &  32.69168	& 32.94095	& 32.7415 \\
      240  &	0.13637831 &	32.73079	& 32.79243	& 32.7431 \\ 
      480  &	0.06820844 &	32.74005	& 32.75542	& 32.7431 \\
      960  &	0.03410665 &	32.74238	& 32.74622	& 32.7431 \\
      1920 &	0.01705362 &	32.74295	& 32.74391	& 32.7431 \\
      \hline
    \end{tabular}}
  \label{RopeLengthOf31}
\end{table}

Similar tests have been performed for the next torus knots: $5_1$, $7_1$ and $9_1$. 

\begin{table}
\caption{The raw polygonal, inscribed and weighted average
    ropelength of the $5_1$ knot tightened by SONO.}
    {\begin{tabular}{|c|c|c|c|c|} \hline
      $n$ & $dl$ & $L_p$ & $L_c$ &$L_a=(4 L_p+L_c)/5$ \\
      \hline 
180  &	0.26182883	& 47.12919	& 47.45869	& 47.20 \\
360  &	0.13105733	& 47.18064	& 47.26225	& 47.20 \\ 
720  &	0.06554618	& 47.19325	& 47.21360	& 47.20 \\
1440 &	0.03277605	& 47.19751	& 47.20261	& 47.20 \\
\hline
    \end{tabular}}
  \label{RopeLengthOf51}
\end{table}

The applicability of the weighted average ropelength to other knots
requires further analysis.

\begin{table}
  \caption{The raw polygonal, inscribed and weighted average
    ropelength of the $7_1$ knot tightened by SONO.}
    {\begin{tabular}{|c|c|c|c|c|} \hline
      $n$ & $dl$ & $L_p$ & $L_c$ &$L_a=(4 L_p+L_c)/5$ \\
      \hline
154	  & 0.39748	& 61.21150	& 62.21861	& 61.41 \\ 
308	  & 0.19921	& 61.35573	& 61.60250	& 61.40 \\
616	  & 0.09966	& 61.39211	& 61.45352	& 61.40 \\
1232	& 0.04984	& 61.40131	& 61.41663	& 61.40 \\
      \hline
    \end{tabular}}
  \label{RopeLengthOf71}
\end{table}

\begin{table}
  \caption{The raw polygonal, inscribed and weighted average
    ropelength of the $9_1$ knot tightened by SONO.}
    {\begin{tabular}{|c|c|c|c|c|} 
	\hline
	$n$ & $dl$ & $L_p$ & $L_c$ &$L_a=(4 L_p+L_c)/5$ \\
	\hline
188	  & 0.40140	& 75.46436	& 76.72711	& 75.72 \\ 
376	  & 0.20118	& 75.64443	& 75.95430	& 75.71 \\
752	  & 0.10066	& 75.69522	& 75.77243	& 75.71 \\
1504	& 0.05034	& 75.70902	& 75.72832	& 75.71 \\
      \hline
    \end{tabular}}
  \label{RopeLengthOf91}
\end{table}

\section{Discussion}

Numerical experiments carried out by the SONO algorithm provide us
with polygonal knots which can be seen as skeletons of tight
knots tied on a corrugated rope.  The vertices of the polygonal knots
are the centers of the spherical segments of the corrugated tube.  As
we demonstrated, one can place a smooth
tube of a smaller radius inside the corrugated tube to obtain 
a smooth knot tied within a smooth tube. 
We know that $L_c$, the bound on its ropelength, is larger
than the ropelength of the ideal knot.  In observing the ropelength of
the inscribed knots, we see that it converges with the increasing
number of vertices to a value $L_\infty$, which can be seen as an
upper bound for the ropelength of an ideal knot. Determining the
value of
$L_\infty$ is a subtle problem and shall not be discussed here;
however, by combining the results we obtained for finite $n$ 
with the inscribing algorithm, we are able to find an upper bound for the
ropelength of a few ideal knots. For the trefoil knot, the bound equals
32.744, for the $5_1$ knot it equals 47.203, for the $7_1$ knot it
equals 61.417 and for the $9_1$ knot it is 75.728. These are the
most precise estimations of the upper bounds obtained so far.

As said above, the $L_c$ values we find analyzing the polygonal knots
delivered by SONO can be seen as upper bounds for the ropelength of
their ideal smooth conformations. But what about the weighted average
values $L_a$? To see their utility we must adopt the less
rigorous point of view of an experimental physicist. The coordinates of
the polygonal knots delivered by SONO can be seen as results of
measurements carried out on knots tied on the corrugated
rope. Analyzing the experimental data, one finds that the
knot length calculated according to the raw polygonal length formula
systematically increases with $n$ while the length calculated
according to the inscribed arcs formula systematically decreases with
$n$. Plotting the values together, one notices that they
converge to a common value: $L_c$ from below, $L_p$ from above. One
also finds that, located between the two plots, the plot of their
weighted average is almost flat and horizontal. By looking at the $L_a$
values presented in the tables, one can clearly see that $L_a$ provides
a good estimate of the ropelength, even at a small number of
vertices.

\section{Acknowledgments}

We thank Maciej Oszwaldowski for indicating the possibility of an
analytical analysis of the weighted average algorithm. 
Pieranski and Baranska were supported under project PB 62 204/04 BW.
Rawdon was supported by NSF Grant No.~0296098.

\bibliographystyle{plain}
\bibliography{wonderbibv9}

\end{document}